\documentclass[
    ,final            
  ]
  {aipproc}

\usepackage{feynarts,amsmath,amssymb,colordvi,gnuplotcolors,psfrag}

\layoutstyle{6x9}

\newcommand{\gev}{\text{GeV}}
\newcommand{\tev}{\text{TeV}}
\newcommand{\MSF}{ M_{\text{Sf.}}}
\newcommand{\tb}{\tan\beta}
\newcommand{\TE}[1]{\cdot 10^{#1}}
\newcommand{\E}[1]{10^{#1}}
\newcommand{\MSSM}{{\text{MSSM}}}
\newcommand{\SM}{{\text{SM}}}
\newcommand{\LO}{{\text{LO}}}
\newcommand{\NLO}{{\text{NLO}}}
\newcommand{\NC}{{\text{NC}}}
\newcommand{\CC}{{\text{CC}}}
\newcommand{\XC}{{\text{XC}}}

\begin{document}

\title{
Towards a Realistic MSSM Prediction for
Neutrino-nucleon Deep-inelastic Scattering
\footnote{
Talk given by O. Brein at {\em SUSY06, the 14th 
International Conference on Supersymmetry
and the Unification of Fundamental Interactions}, Irvine,
California, 12--17 June 2006.}
}


\author{Oliver Brein}{
  address={Institute for Particle Physics Phenomenology,
University of Durham, DH1 3LE,\\ 
Durham, United Kingdom}
}

\author{Wolfgang Hollik}{
  address={Max-Planck-Institut f\"ur Physik,
     F\"ohringer Ring 6, D-80805 M\"unchen, Germany}
}

\author{Benjamin Koch}{
  address={Institut f\"ur Theoretische Physik,
Johann Wolfgang Goethe-Universit\"at,\\
 D-60054 Frankfurt am Main, Germany}
}

\begin{abstract}
We discuss the radiative corrections to charged and neutral current
deep-inelastic neutrino--nucleon scattering in the minimal supersymmetric
standard model (MSSM).
In particular, we study 
deviations, $\delta R^{\nu (\bar\nu)}$, from the Standard Model 
prediction for the ratios of neutral- to charged-current 
cross sections, 
taking into account all
sources for deviations in the MSSM,
i.e.
different contributions from virtual Higgs bosons
and virtual superpartners.
Our calculation includes
the full $q^2$ dependence of the one-loop amplitudes, 
parton distribution functions
and 
a NuTeV-inspired process kinematics.
We present results of a scan of $\delta R^{\nu (\bar\nu)}$
over the relevant MSSM parameter space.
\end{abstract}

\maketitle
{\flushright \vspace*{-11cm}IPPP/06/74\\
DCPT/06/148\\
MPP-2006-139\\
\vspace*{9cm}}

\section{Introduction}
\begin{figure}[t]
{
\unitlength=0.3mm%
\begin{picture}(300,10)(0,0)
\begin{footnotesize}

\begin{feynartspicture}(290,70)(3,1)
\FALabel(44.,91.96)[]{}
\FADiagram{}
\FAProp(0.,15.)(10.,14.)(0.,){/Straight}{1}
\FALabel(0.84577,13.4377)[t]{$\nu_\mu$}
\FAProp(0.,5.)(10.,6.)(0.,){/Straight}{1}
\FALabel(1.15423,4.43769)[t]{{$d,\bar{u}, s,\bar{c}$}}
\FAProp(20.,15.)(10.,14.)(0.,){/Straight}{-1}
\FALabel(19.8458,15.5623)[b]{$\mu$}
\FAProp(20.,5.)(10.,6.)(0.,){/Straight}{-1}
\FALabel(20.1542,5.8)[b]{{$u,\bar{d}, c, \bar{s}$}}
\FAProp(10.,14.)(10.,6.)(0.,){/Sine}{-1}
\FALabel(8.93,10.)[r]{$W$}
\FAVert(10.,14.){0}
\FAVert(10.,6.){0}

\FADiagram{}

\FADiagram{}
\FAProp(0.,15.)(10.,14.)(0.,){/Straight}{1}
\FALabel(0.84577,13.4377)[t]{$\nu_\mu$}
\FAProp(0.,5.)(10.,6.)(0.,){/Straight}{1}
\FALabel(2.15423,4.43769)[t]{$u,\;d,\;s,\;c$}
\FAProp(20.,15.)(10.,14.)(0.,){/Straight}{-1}
\FALabel(19.,15.5623)[b]{$\nu_\mu$}
\FAProp(20.,5.)(10.,6.)(0.,){/Straight}{-1}
\FALabel(20.7542,5.8)[b]{$u,\;d,\;s,\;c$}
\FAProp(10.,14.)(10.,6.)(0.,){/Sine}{0}
\FALabel(8.93,10.)[r]{$Z$}
\FAVert(10.,14.){0}
\FAVert(10.,6.){0}
\end{feynartspicture}
\end{footnotesize}
\end{picture}
}
\caption{\label{born}
Tree-level Feynman graphs for neutral-current and charged-current
scattering of muon-neutrino and quark.}
\end{figure}
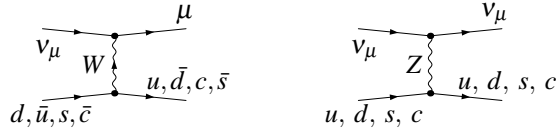

In the Standard Model (SM), neutral (NC) and charged current (CC) neutrino--nucleon
scattering are described in leading order by $t$-channel $W$ and $Z$ exchange, respectively
(see Fig.~\ref{born}).
At the NuTeV experiment, $\nu_\mu$ and $\bar\nu_\mu$ beams of a mean energy
of 125 GeV were scattered
off a target detector and the ratios
$R^\nu = {\sigma_\NC^\nu}/{\sigma_\CC^\nu}$ and
$R^{\bar\nu} = {\sigma_\NC^{\bar\nu}}/{\sigma_\CC^{\bar\nu}}$
were measured.
The NuTeV collaboration also provided a determination of the on-shell weak
mixing angle~\cite{nutev02},
$
\sin^2\theta_w^{\text{\tiny on-shell}} = 0.2277 \pm 0.0013 (stat.)
\pm 0.0009 (syst.)\;.
$
This value is about $3\sigma$ below the value derived from the residual set
of precision observables~\cite{sinthetaSM}.
The fixed-collision-energy ratios 
$R^\nu$ and $R^{\bar\nu}$ were not directly accessible at NuTeV.
More precisely, ratios of counting rates, 
i.e. number of NC-like neutrino events over 
number of CC-like neutrino events
and likewise for anti-neutrino events,
$R^\nu_{\text{exp}}, R^{\bar\nu}_{\text{exp}}$, were measured
and compared to SM predictions, 
$R^\nu_{exp}(SM)$, $R^{\bar{\nu}}_{exp}(SM)$,
obtained by a detailed Monte Carlo (MC) simulation.
These predictions differ from the 
measured values by
{\cite{nutev-DeltaRnu}} 
\begin{align}
\label{deltaRnu-nubar}
\Delta R^{\nu (\bar\nu)} & = R^{\nu(\bar\nu)}_{\text{exp}} 
	- R^{\nu (\bar\nu)}_{\text{exp}}(SM)
= -0.0032 \pm 0.0013 \;(-0.0016 \pm 0.0028)\;.
\end{align}
Basically, there are three types of explanations of the observed anomaly.\\
(a) {\em There is no signal.} The result might be a statistical fluctuation
or theoretical errors may have been underestimated.
This line of thought lead to re-analyses of the electroweak 
radiative corrections to neutrino--nucleon 
deep-inelastic scattering (DIS)\cite{DDH}, 
which have been originally calculated almost 20 years ago
\cite{bardin-dokuchaeva}.\\
(b) {\em The apparent signal is due to neglected but relevant SM effects.}
An asymmetry between the distribution of the strange quark 
and its corresponding anti-quark in the nucleon ($s \not= \bar s$)
could account for part of the deviation.
The result of the measurement of the weak mixing angle could
also be influenced by 
a violation of the usually assumed isospin symmetry ($u_p \not= d_n$),
or nuclear effects which have not been taken into account. 
For a list of other sources see Refs.~\cite{davidson,sm-explanations} and
references therein.\\
(c) {\em The signal is due to new physics.}
Many suggestions have been made, like effects from 
modified gauge boson interactions (e.g. in extra dimensions), 
non-renormalizable operators,
leptoquarks (e.g. $R$-parity violating supersymmetry) 
and supersymmetric loop effects (e.g. in the MSSM), just to name a few of them (see
\cite{davidson,shufang} and
references therein for an overview).

Here we consider the effects of 
the radiative corrections to 
neutrino--nucleon DIS in the MSSM.
Although the NuTeV anomaly is not  settled as yet,
it is interesting to check how far the MSSM could account for such a
deviation.
Two earlier studies~\cite{davidson,shufang}
treat the loop effects in the limit of zero-momentum transfer
of the neutrino to the hadron, neglecting kinematical cuts and 
potential effects from the parton distribution functions.
They conclude
that the radiative corrections in the MSSM cannot be made
responsible for the NuTeV anomaly, owing to the wrong sign.

Our calculation \cite{our-calc} includes various kinematical effects. 
In particular,
we include the full $q^2$ dependence of the one-loop amplitudes,
evaluate hadronic cross sections using 
parton distribution functions and 
cuts on the hadronic energy in the final state 
($20\,\gev < E_{\text{had.}} < 180\,\gev$)
at the NuTeV mean neutrino beam energy of 125 GeV.
Moreover, we perform  
a thorough parameter scan for the 
radiative corrections $\delta R^{\nu(\bar\nu)}$
over the relevant MSSM parameter space.

\section{MSSM radiative corrections to $\nu_\mu N$ DIS}

The difference between the MSSM and SM predictions,
$\delta R^n = R^n_\MSSM - R^n_\SM$ with
$R^n = {\sigma^n_\NC}/{\sigma^n_\CC}$,
with
$
(\sigma^n_\XC)_\NLO  = (\sigma^n_\XC)_\LO + \delta\sigma^n_\XC 
\; (\text{X = N,C} ; n = \nu,\bar\nu)
$,
can be expanded as follows,
\begin{align*}
\delta R^n & = \left(\frac{\sigma^n_\NC}{\sigma^n_\CC}\right)_\LO 
 \left(\frac{(\delta\sigma^n_\NC)_\MSSM-(\delta\sigma^n_\NC)_\SM}{(\sigma^n_\NC)_\LO}
-\frac{(\delta\sigma^n_\CC)_\MSSM-(\delta\sigma^n_\CC)_\SM}{(\sigma^n_\CC)_\LO}
\right) .
\end{align*}
Thus, only leading-order (LO) cross-sections or
differences between MSSM and SM radiative corrections appear.
$R$-parity conservation in the MSSM makes
the Born cross section the same as in the SM (up to a negligible 
extra contribution involving the charged Higgs boson).
Consequently, 
contributions from real photon emission
and all SM-like radiative corrections without virtual Higgs bosons
are equal in 
both models.
Therefore, the differences between the MSSM and SM 
one-loop radiative corrections
boils down
to the genuine superpartner (SP) loops and the difference between 
the Higgs-sector contributions, i.e.\ they are schematically given by
\begin{align}
\label{delR-schema}
\delta\sigma_\MSSM - \delta\sigma_\SM& \propto 
\big(\; [\text{SP loops}]
+ [\text{Higgs graphs MSSM}-\text{Higgs graphs SM}] \;\big) \;.
\end{align}
The second term in Eq.~(\ref{delR-schema}) vanishes
when the MSSM Higgs sector is SM-like, which is the case for 
a $CP$-odd Higgs mass $m_A \gtrsim 250\,\gev$.
The superpartner-loop contributions for CC (anti-)neutrino--quark scattering
consist of $W$-boson selfenergy insertions, loop contributions to 
the $Wq\bar q'$- and $W\nu_\mu \mu^- (W\bar\nu_\mu \mu^+)$-vertex, and
box-graphs with double-exchange of neutralinos and charginos.
Analogous contributions appear in the NC case with $W$ replaced by
$Z$ and $\mu^-(\mu^+)$ by $\nu_\mu(\bar\nu_\mu)$. 
Additionally, there are SP-loop contributions to the 
photon--$Z$ mixing selfenergy.
The difference between MSSM and SM Higgs sector contributions
to the CC (NC) processes
consists of the 
difference between all $W$($Z$) selfenergy contributions
which contain at least one virtual physical Higgs boson.

\begin{figure}
{\setlength{\unitlength}{1cm}
\begin{picture}(15,8)(0,-.2)
\psfrag{XT}[c][l][2][90]{$X_t\;[\tev]$}
\psfrag{MNEU1}[l][c][2][90]{$m_{\chi^0_1}\;[\gev]$}

\put(-1.5,3){\resizebox*{.43\width}{.43\height}{\includegraphics*{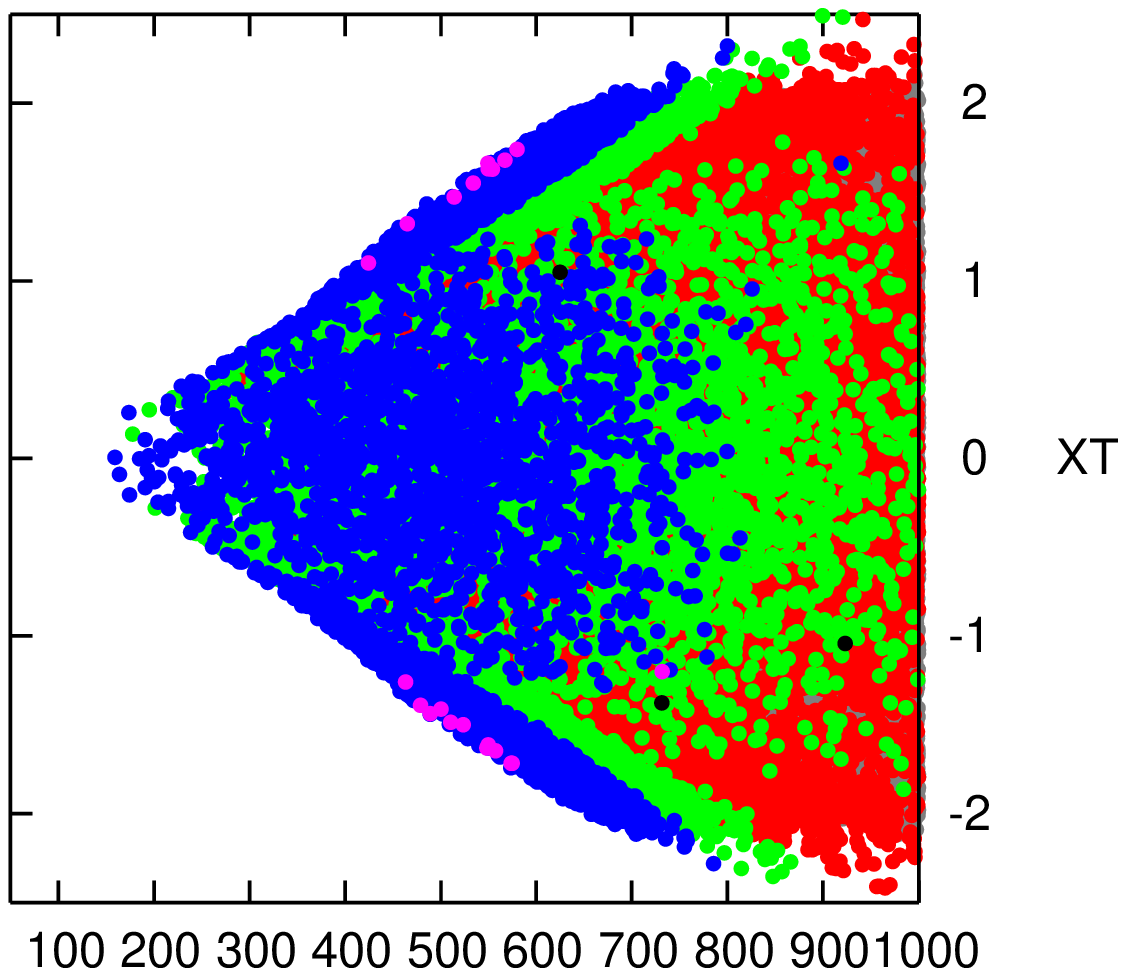}}}
\put(.6,8.4){(a)}
\put(.6,7.9){$\delta R^{\bar\nu}$}
\put(-1.5,-1.4){\resizebox*{.43\width}{.43\height}{\includegraphics*{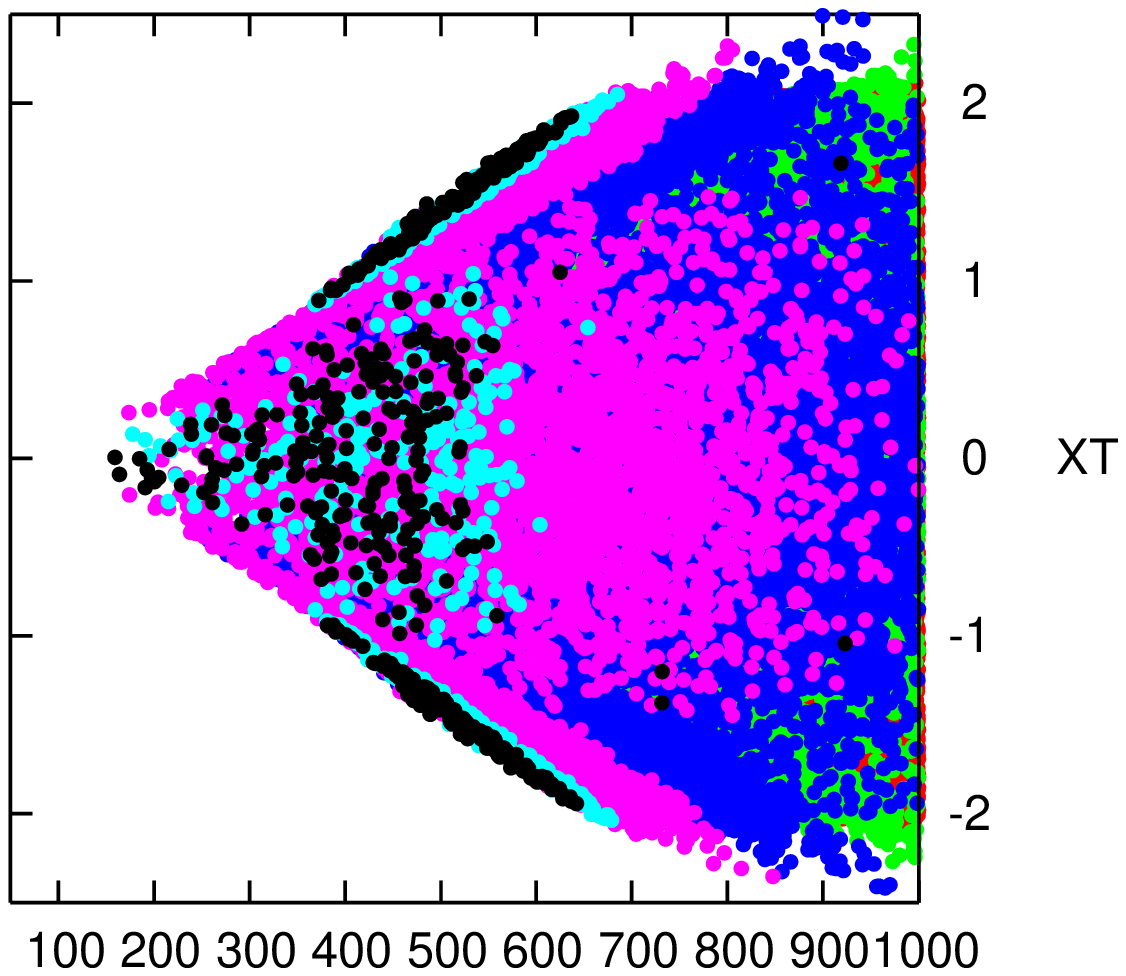}}}
\put(.6,4){(b)}
\put(.6,3.5){$\delta R^\nu$}

\put(10.,0){\resizebox*{0.95\width}{0.95\height}{\includegraphics*{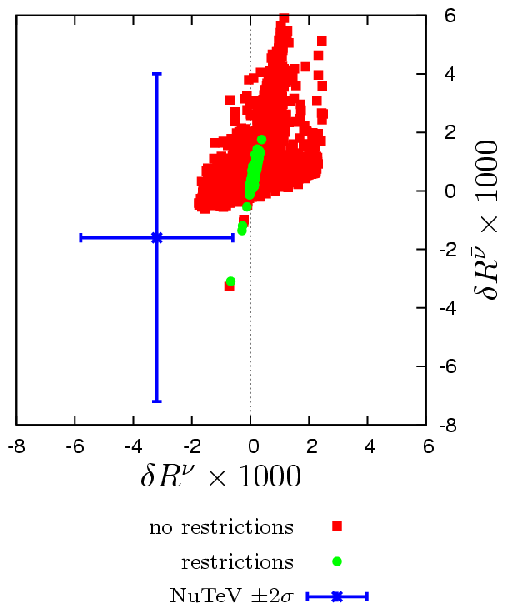}}}
\put(10.25,5.25){(d)}

\put(3.5,0.4){\resizebox*{.43\width}{.43\height}{\includegraphics*{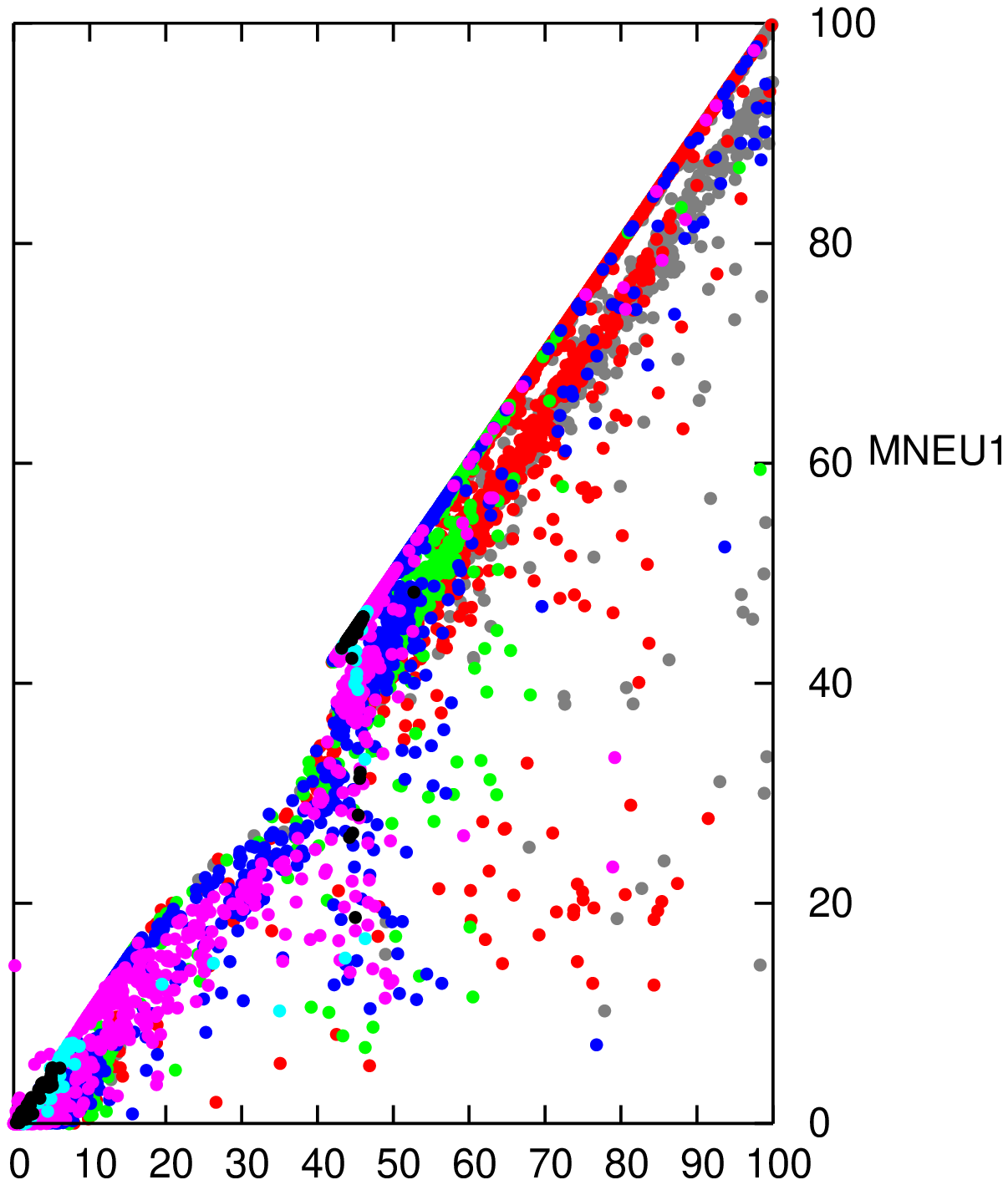}}}
\put(5.5,8.4){(c)}
\put(5.5,7.9){$-\delta R^{\nu}$}

\put(6.1,2.25){\footnotesize $\delta R^{\nu(\bar\nu)}$}
\put(5.6,1.95){\GNUPlotI{$\scriptstyle 0 - 1\TE{-5}$}}
\put(5.6,1.6){\GNUPlotA{$\scriptstyle 1\TE{-5} - 5\TE{-5}$}}
\put(5.6,1.25){\GNUPlotB{$\scriptstyle 5\TE{-5} - 1\TE{-4}$}}
\put(5.6,0.9){\GNUPlotC{$\scriptstyle 1\TE{-4} - 3\TE{-4}$}}
\put(5.6,0.45){\GNUPlotD{$\scriptstyle 3\TE{-4} - 8\TE{-4}$}}
\put(5.6,0.1){\GNUPlotE{$\scriptstyle 8\TE{-4} - 1\TE{-3}$}}
\put(5.6,-0.25){\GNUPlotG{$\scriptstyle > 1\TE{-3}$}}

\put(2,-.2){\small $\MSF\;[\gev]$}

\put(7.8,2.4){\small $m_{\chi^\pm_1}\;[\gev]$}

\end{picture}
}
\caption{\label{drnu}
Results of parameter scan I (see text) for (a) $\delta R^\nu$ and
(b) $\delta R^{\bar\nu}$ in the $\MSF$--$X_t$--plane. Panel (c)  
shows $-\delta R^\nu$ of all results with $\delta R^\nu < 0$ 
of scan II in the $m_{\chi^\pm_1}$--$m_{\chi^0_1}$--plane. 
In panels (a), (b) and (c)
points with larger values of are plotted on top of points with smaller
values.
Panel (d) shows the results of scan I
and II in the $\delta R^\nu$-$\delta R^{\bar\nu}$ plane together
with the NuTeV $2\sigma$ intervals 
for $\Delta  R^\nu$ and $\Delta  R^{\bar\nu}$.
}
\end{figure}

\section{MSSM parameter scan for $\delta R^{\nu(\bar\nu)}$}

We perform two different parameter scans over the following 
region of MSSM parameter space:
$10\,\gev  \leq M_1, M_2, M_3, \MSF, m_{A^0}\leq 1\,\tev$, 
$1 \leq \tb \leq 50$, $-2\,\tev \leq \mu, A_t, A_b, A_\tau \leq 2\,\tev$,
where $M_1, M_2, M_3$ are gaugino mass parameters,
$\MSF$ is a common sfermion mass scale,
$\tb$ is the ratio of the two Higgs vacuum expectation values in the
MSSM,
$\mu$ is the supersymmetric Higgs mass term,
$A_t, A_b$ and $A_\tau$ are soft-breaking trilinear couplings.
For both scans,
we use the ``adaptive scanning'' technique \cite{adaptive-scan}
to scan the parameter space 
with an emphasis on regions where $\delta R^\nu, \delta R^{\bar\nu}$
are close to the observed deviations from the SM (Eq.~\ref{deltaRnu-nubar}).
In scan I,  
we neglect all points in parameter space
which violate mass exclusion limits for Higgs bosons
or for superpartners or the $\Delta\rho$-constraint on sfermion mixing.
Fig.~\ref{drnu}(a) and \ref{drnu}(b) show the resulting values  
for $\delta R^{\nu}$ and $\delta R^{\bar\nu}$, respectively, 
in the $\MSF$--$X_t$--plane ($X_t = A_t - \mu/\tb$).
Particularly large radiative corrections occur
for large stop-mixing ($|X_t| \approx 3 \MSF$).
Interestingly, the only negative values we obtain for 
$\delta R^{\nu}$  and $\delta R^{\bar\nu}$ have 
magnitudes below $\E{-4}$ and are not shown 
in Figs.~\ref{drnu}(a) and \ref{drnu}(b).
In scan II, we allow the parameter points to violate 
the constraints applied in scan I in order to
see whether parameter regions, which could 
explain the NuTeV measurement, exist at all.
Indeed, we find such points.
As an example, Fig.~\ref{drnu}(c) shows only parameter points with
$\delta R^{\nu} < 0$
resulting from scan II
in the plane of the lightest neutralino and chargino mass
($m_{\chi^0_1}, m_{\chi^\pm_1}$), 
which proves to be decisive for the sign of 
$\delta R^{\nu(\bar\nu)}$. 
Negative values with a magnitude $> \E{-3}$
only appear for $m_{\chi^\pm_1} \lesssim 60\,\gev$.
Thus, already the LEP-constraint on the chargino mass, 
$m_{\chi^\pm_1} > 94\,\gev$, excludes such MSSM scenarios.
For 
heavier charginos or neutralinos
than  shown in Fig.~\ref{drnu}(c) 
$|\delta R^\nu|$ always stays below $\E{-3}$.
Fig.~\ref{drnu}(d) relates the result of the two scans to the
NuTeV measurements of $\Delta R^{\nu(\bar\nu)}$. 
While scenarios with compatible
values for $\delta R^\nu$ and $\delta R^{\bar\nu}$ are possible
if no restrictions are taken into account, the possible predictions 
for $\delta R^\nu$ lie outside the $2\sigma$ range if 
restrictions are applied.

\section{Summary}
The NuTeV measurement of $\sin^2\theta_w$ is 
intriguing but has to be further established.
Especially, confirmation by other experiments is desirable.
We calculated the MSSM radiative corrections to the cross 
section ratios $R^{\nu (\bar\nu)}$ including several kinematic 
effects.
It seems,
MSSM loop effects 
do not provide a viable explanation of the deviation
observed by NuTeV.
The size of the deviation could be of the right order, but
it either appears with the wrong sign or violates other electroweak
constraints.
In any case, 
interesting restrictions on the
MSSM parameters can be obtained.
Combined with the NLO prediction in the SM,
our calculation can easily provide a simulation tool for 
neutrino-nucleon scattering in the MSSM 
at the same level of accuracy.

\bibliographystyle{aipproc}   

\begin{thebibliography}{9}

\bibitem{nutev02} 
G.~P.~Zeller et al.\  [NuTeV Collaboration],
Phys.\ Rev.\ Lett.\  {\bf 88} (2002) 091802. 
[Erratum-ibid.\  {\bf 90} (2003) 239902].

\bibitem{sinthetaSM} 
The LEP Collaborations, the LEP Electroweak Working Group and the
SLD Heavy Flavor and Electroweak Working Groups,
Phys.\ Rept.\  {\bf 427} (2006) 257.


\bibitem{nutev-DeltaRnu}
G.~P.~Zeller et al.\  [NuTeV Collaboration],
hep-ex/0207052.

\bibitem{DDH}
K.-P.~O.~Diener, S.~Dittmaier and W.~Hollik,
Phys.\ Rev.\ D {\bf 69} (2004) 073005,
ibid. {\bf 72} (2005) 093002;


A.~B.~Arbuzov, D.~Y.~Bardin and L.~V.~Kalinovskaya,
JHEP {\bf 0506} (2005) 078.

\bibitem{bardin-dokuchaeva}
D.~Y.~Bardin and V.~A.~Dokuchaeva,
JINR-E2-86-260

\bibitem{davidson}
S.~Davidson et al.,
JHEP {\bf 0202} (2002) 037.

\bibitem{sm-explanations}
S.~A.~Kulagin,
Phys.\ Rev.\ D {\bf 67} (2003) 091301;
K.~S.~McFarland and S.~O.~Moch,
hep-ph/0306052;
S.~Kretzer et al.,
Phys.\ Rev.\ Lett.\  {\bf 93} (2004) 041802.

\bibitem{shufang}
A.~Kurylov, M.~J.~Ramsey-Musolf and S.~Su,
Nucl.\ Phys.\ B {\bf 667} (2003) 321.

\bibitem{our-calc}
O.~Brein, B.~Koch and W.~Hollik, Proc. Internat. Conf.
on Linear Colliders, LCWS 04, Paris, France, April 19-23,
hep-ph/0408331.

\bibitem{adaptive-scan}
O.~Brein,
Comput.\ Phys.\ Commun.\  {\bf 170}, 42 (2005).

\end{thebibliography}

\end{document}